\documentclass[preprint,a4paper,english,superscriptaddress,aps]{revtex4}
\usepackage[dvips]{graphicx}
\usepackage[all]{xy}

\usepackage[english]{babel}
\usepackage{amsmath}
\newcommand{\I}{\mathrm{i}}
\newcommand{\e}{\mathrm{e}}

\begin{document}

\title{Landau Analog Levels for Dipoles in the Noncommutative Space and Phase Space}

\author{L. R. Ribeiro}
\email{lrr,passos,jroberto,furtado@fisica.ufpb.br}
\affiliation{{ Departamento de
F\'{\i}sica, Universidade Federal da Para\'\i ba, Caixa Postal 5008, 58051-970,
Jo\~ao Pessoa, PB, Brasil}}

\author{E. Passos}
\affiliation{{ Departamento de
F\'{\i}sica, Universidade Federal da Para\'\i ba, Caixa Postal 5008, 58051-970,
Jo\~ao Pessoa, PB, Brasil}}

\author{C. Furtado}
\affiliation{{ Departamento de
F\'{\i}sica, Universidade Federal da Para\'\i ba, Caixa Postal 5008, 58051-970,
Jo\~ao Pessoa, PB, Brasil}}

\author{J. R. Nascimento}
\affiliation{{ Departamento de
F\'{\i}sica, Universidade Federal da Para\'\i ba, Caixa Postal 5008, 58051-970,
Jo\~ao Pessoa, PB, Brasil}}
\affiliation{Instituto de F\'\i sica, Universidade de S\~ao Paulo\\
Caixa Postal 66318, 05315-970, S\~ao Paulo, SP, Brazil}

\begin{abstract}
In the present contribution we investigate the Landau analog energy quantization for neutral particles, that possesses a nonzero permanent magnetic and electric dipole moments, in the presence of an homogeneous electric and magnetic external fields in the context of the noncommutative quantum mechanics. Also, we analyze
the Landau--Aharonov--Casher and Landau--He--McKellar--Wilkens quantization due to noncommutative quantum dynamics of  magnetic and electric dipoles in the presence of an external electric and magnetic fields and the energy spectrum and the eigenfunctions are obtained.
Furthermore, we have analyzed Landau quantization analogs in the noncommutative phase space, and we obtain also the energy spectrum and the eigenfunctions in this context.
\end{abstract}
\maketitle

\section{Introduction}
Several topological and geometrical effects may be realized studying the quantum dynamics of charged and neutral particles in presence of electric and magnetic fields. In 1959, Aharonov and Bohm demonstrated that electromagnetic fields affect the state of matter even in spatial regions where the field is zero \cite{aha}. This effect occurs due to presence of the vector potential in the region where the particle moves. In other words, in a multiply-connected region of space where there's no fields, in quantum mechanics, the physical properties of the system still depend on the potential, in contrast of the classical physics. In the Aharonov--Bohm effect, a quantum charged particle circulating around a magnetic flux line acquires a quantum topological non-dispersive phase in its wave function. This effect was observed experimentally by Chambers \cite{prl:cham,pes1}. 

Years later, in 1984, Aharonov and Casher demonstrated that the wave function of a neutral particle that possess a non-zero magnetic dipole moment, moving through a non-simply-connected force-free region, is affected by an electric field in a similar way of the Aharonov--Bohm effect \cite{cas}. In the Aharonov--Casher effect, a quantum neutral particle with non-zero magnetic dipole moment, moving around and parallel to a charged wire, accumulates on the wave function a quantum geometrical non-dispersive phase. This effect was observed in a neutron interferometer \cite{cim} and in a neutral atomic Ramsey interferometer \cite{san}. He and McKellar in 1993 \cite{mac}, and Wilkens \cite{wil} independently in 1994, predicted the existence of a quantum phase acquired by the wave function of a neutral particle, that possess a non-zero electric dipole moment, while it is circulating around and parallel to a line of magnetic monopoles. A simple practical experimental setup to test the He--McKellar--Wilken effect, without the inconvenience of magnetic monopoles, was proposed by Wei et al. \cite{wei}. In this setup, a electric dipole moment is induced on the neutral particle by the electric field of a charged wire and a uniform magnetic field is applied crossed to the electric field. Another two experimental schemes for the He--McKellar--Wilkens phase are proposed by Dowling et al. \cite{franson}, as well as a unified description of all three phenomena and have studied a new effect the dual Aharonov-Bohm effect. The dual Aharonov-Bohm phase can be calculated in the quantum dynamics of a magnetic monopole in the presence of an electric solenoid\cite{furtadodual}.

In 1930, Landau showed that a charged particle moving in an homogeneous magnetic field presents quantized energy levels \cite{landau}. The Landau levels rises a remarkable interest to describe several problems in physics, e.g. quantum Hall effect \cite{prange}, different two-dimensional surfaces \cite{comtet,dunne}, anyons excitations in a rotating Bose--Einstein condensate \cite{paredes1,paredes2}, and others like analog levels for dipoles. Ericsson and Sj\"oqvist developed a analog of Landau quantization for neutral particles in presence of a external electric field \cite{ericsson}. The idea is based on the Aharonov--Casher effect in which neutral particles may interact with an electric field via a non-zero magnetic dipole moment. In the same way, we developed a analog Landau quantization for neutral particles, that possess a non-zero electric dipole moment, making use of the He--McKellar--Wilkens effect \cite{lrr:pla1}. To solve the problem of magnetic monopoles, we proposed the study of a Landau analog quantization in quantum dynamics of an induced electric dipole in the presence of crossed electric and magnetic fields \cite{lrr:pla2}.

Recently, noncommutative space theories, motivated by string theory \cite{sei, sei2}, has attracted interest in several areas of physics \cite{nekra}; e.g. quantum gravity \cite{moffat}, M-theory \cite{con}, and quantum Hall effect \cite{sus,jela,basu}. In quantum mechanics, a great number of problems have been investigated in the case of noncommutative space\cite{hovarti} and phase-space. Some important results obtained are related to geometric phases, such as the Aharonov--Bohm effect \cite{1,2,3,4,5}, the Aharonov--Casher effect \cite{mirza,kangepjc}, the Berry quantum phase \cite{alavi,ghos}, Landau levels\cite{hvarti2,gamboa2,hovarti3} and others that involve dynamics of dipoles \cite{cal}. In a recent paper, we analyzed the quantum geometrical phase effect for a quantum neutral particle with permanent magnetic and electric dipole moments in the presence of external magnetic and electric fields, proposed by Anandan \cite{prlan},  in the noncommutative space and phase space quantum mechanics context \cite{lrr:pra}.

The aim of this work is the study of the Landau analogs energy levels for neutral particles, that possesses nonzero magnetic and electric dipole moments, in the presence of homogeneous electric and magnetic external fields in the context of noncommutative quantum mechanics. We calculate the corrections in analog Landau energy levels due to the noncommutativity of space and phase space coordinates. 

The paper is organized as follows: In the section \ref{2} we will presents a review of the standard Landau quantization for a charged particle moving in the homogeneous external magnetic field \cite{landau}. In the section \ref{3} we make a short review of the analog Landau quantization for magnetic and electric dipoles in the presence of external magnetic and electric  fields \cite{ericsson,lrr:pla1}. In the section \ref{4} a general overview of the noncommutative quantum mechanics is presented \cite{Bopp}. In the sections \ref{5}, \ref{6}, \ref{7} and \ref{8} we investigate the Landau analog effects in the noncommutative space and phase space. Finally, in the section \ref{9} we present the conclusions.

\section{Charged Particle in a Homogeneous Magnetic Field}\label{2}
Considering a charged particle, with charge $-e$, moving in the plane $x$-$y$ and submitted to a external homogeneous magnetic field oriented in the $z$-axis direction, $\vec{B}=B_0\hat{\e}_z$, we may obtain quantized energy levels for this particle. These energy levels are called Landau levels \cite{landau}. The Hamiltonian of the system is given by (we used the natural units system $\hbar=c=1$)
\begin{equation}
	H=\frac{1}{2m}\left(\hat{p}-e\vec{A}\right)^2\;,
	\label{eqc:1}
\end{equation}
where $\vec{A}$ is the vector potential, $\vec{B}=\nabla\times\vec{A}$, and $\hat{p}$ is the linear momentum, $\hat{p}=-\I\nabla$. Choosing the symmetric gauge
\begin{equation}
	\vec{A}=\dfrac{B_0}{2}(-y,x,0)=\dfrac{B_0}{2}r\hat{\e}_\phi\;,
	\label{eqc:2}
\end{equation}
where in cylindrical coordinates $r^2=x^2+y^2$ and $\hat{e}_\phi$ is the unitary vector oriented in $\phi$-direction. Hence, from the Eq.(\ref{eqc:1}) we can write the Schr\"odinger equation, making use of the cylindrical symmetry of the system, in the form
\begin{equation}
	-\frac{1}{2m}\left[\frac1r\frac{\partial}{\partial r}\left(r\frac{\partial\psi}{\partial r}\right)+\frac{1}{r^2}\frac{\partial^2\psi}{\partial\phi^2}\right]-\frac{\I\omega}{2}\frac{\partial\psi}{\partial\phi}+ \frac{m\omega^2}{8}r^2\psi=\mathcal{E}\psi\;,
	\label{eqc:3}
\end{equation} 
where
\begin{equation}
	\omega=\frac{eB}{m}\;
	\label{eqc:4}
\end{equation}
is the cyclotron frequency. Solving the 	Eq.(\ref{eqc:3}), we find the Landau levels that is given by
\begin{equation}
	\mathcal{E}=\left(n+\frac{|\ell|+\ell+1}{2}\right)\omega\;,
	\label{eqc:5}
\end{equation}
where $\ell$ is a integer number related to the wave function periodicity, in the form
\begin{equation}
	\psi=\e^{\I\ell\phi}R(r)\;,
	\label{eqc:6}
\end{equation}
and $R(r)$ is the radial eigenfunction written as 
\begin{equation}
	R_{n,\ell}(r)=\frac{1}{a^{|\ell|+1}}\left[\frac{(|\ell|+n)!}{2^{|\ell|}n!|\ell|!^2}\right]\e^{-\frac{r^2}{4a^2}}r^{|\ell|}F\left[-n,|\ell|+1,\frac{r^2}{2a^2}\right]\;,
	\label{eqc:7}
\end{equation}
where
\begin{equation}
	a=\sqrt{\frac{1}{m\omega}}\;
	\label{eqc:8}
\end{equation}
is the magnetic length. Here, $F$ is the degenerated hypergeometric function.

\section{Landau Levels Analog for Dipoles}\label{3}
Considering the nonrelativistic limit of a single neutral spin-half particle, with nonzero magnetic and electric dipole moments, moving in an external electromagnetic field \cite{lrr:pra}. In this limit, and neglecting terms of $O(\vec{E}^2)$ and $O(\vec{B}^2)$, the Anandan's Hamiltonian is write as
\begin{equation}
	H=-\frac{1}{2m}\left[\nabla-\I(\vec{\mu}\times\vec{E})+\I(\vec{d}\times\vec{B})\right]^2-\frac{\mu}{2m}\nabla\cdot\vec{E}+ \frac{d}{2m}\nabla\cdot\vec{B}\;,
	\label{eqa:1}
\end{equation}
where $\vec{\mu}$ and $\vec{d}$ are the magnetic and electric dipole moments of the particle; $\vec{B}$ and $\vec{E}$ are the magnetic and electric fields.

Under certain dipole-field configurations, analog effects of the standard Landau quantization occurs. In this sense, we make use of the Aharonov--Casher and He--McKellar--Wilkens effects, in which neutral particles may interact with electric and magnetic fields via dipole moments. First consider the case in which $d$ and $\vec{B}$ vanishes in Eq.(\ref{eqa:1}) and the Landau--Aharonov--Casher Hamiltonian is given by
\begin{equation}
	H=\frac{1}{2m}\left[\vec{p}-\mu\vec{A}_{AC}\right]^2-\frac{\mu}{2m}\nabla\cdot\vec{E}\;,
	\label{eqa:2}
\end{equation}
where the effective vector potential is given by
\begin{equation}
	\vec{A}_{AC}=\vec{n}\times\vec{E}\;,\qquad\vec{n}=\frac{\vec{\mu}}{|\vec{\mu}|}\;,
	\label{eqa:3}
\end{equation}
and $\vec{n}$ is the unitary vector oriented in the dipole direction, so $\vec{\mu}=\mu\vec{n}$. We may define the associated field strength
\begin{equation}
	\vec{B}_{AC}=\nabla\times\vec{A}_{AC}\;.
	\label{eqa:4}
\end{equation}

The precise field-dipole configuration under which the Landau--Aharonov--Casher occurs was demonstrated by Ericsson and Sj\"oqvist \cite{ericsson}. The conditions are vanishing torque on the dipole, electrostatics $\partial_t\vec{E}=0$, $\nabla\times\vec{E}=0$ and $B_{AC}$ is uniform. Choosing $\vec{n}$ parallel to $z$-axis $\vec{n}=\hat{e}_z$, the two first conditions are fulfilled if the electric field $\vec{E}$ is smooth and $E_z=0$, and the particle moves in the $x$-$y$ plane. So, $\nabla\cdot\vec{E}$ reduces to Gauss's law and $\vec{B}_{AC}=\rho\hat{\mathrm{e}}_z$, where $\rho$ is a uniform volume charge density, fulfilling the third condition.

Now choosing the field configuration for the symmetric gauge as 
\begin{eqnarray}
	\vec{E}&=&\frac{\rho}{2}r\hat{\mathrm{e}}_r\;\label{eqa:5},
\end{eqnarray}
and we obtain the following effective potential
\begin{eqnarray}
	\vec{A}_{AC}&=&\frac{\rho}{2}r\hat{\mathrm{e}}_\phi\;,
\end{eqnarray}
we rewrite the Eq.(\ref{eqa:2}), making use of the cylindrical symmetry, in the form
\begin{equation}
	H=\frac{1}{2m}\left[\vec{p}-\frac{m\omega}{2}r\hat{\mathrm{e}}_\phi\right]^2-\frac{\omega}{2}\;,
	\label{eqa:6}
\end{equation}
where
\begin{equation}
\omega=\omega_{AC}=\frac{\mu\rho}{m}\;
\label{eqa:7}
\end{equation}
is the cyclotron frequency. Therefore, The Landau--Aharonov--Casher energy levels are given by\cite{ericsson}
\begin{equation}
	\mathcal{E}=\left(n+\frac{|\ell|-\ell+1}{2}-\frac{1}{2}\right)\omega_{AC}\;,
	\label{eqa:8}
\end{equation}
where $n=0,1,2\dots$. Therefore, we have energy levels for neutral magnetic polarized particles moving in a electric field in the same way of Landau quantization for charged particles in a homogeneous magnetic field.

In the same way, in the case in which $\vec{\mu}$ and $\vec{E}$ vanishes, we may define another analog effect. Thus, writing the Landau--He--McKellar--Wilkens Hamiltonian as
\begin{equation}
	H=\frac{1}{2m}\left[\vec{p}+d\vec{A}_{HMW}\right]^2+ \frac{d}{2m}\nabla\cdot\vec{B}\;,
	\label{eqa:9}
\end{equation}
where
\begin{equation}
	\vec{A}_{HMW}=\vec{n}\times\vec{B}\;,\qquad\vec{n}=\frac{\vec{d}}{|\vec{d}|}\;,
	\label{eqa:10}
\end{equation} 
and $\vec{n}$ is the unitary vector oriented in the dipole direction, so $\vec{d}=d\vec{n}$. We may define the associated field strength
\begin{equation}
	\vec{B}_{HMW}=\nabla\times\vec{A}_{HMW}\;.
	\label{eqa:11}
\end{equation}
In this case we may also determine the field-dipole configuration under which the Landau--He--McKellar--Wilkens effect occurs. In the same way of the Landau--Aharonov--Casher effect, in this case the torque on the dipole can be vanish, $\partial_t\vec{B}=0$ and $\vec{B}$ must be smooth, and $\vec{B}_{HMW}$ must be uniform. Thus, if $\vec{n}=\hat{e}_z$, $B_z=0$ and the particle moves on the $x$-$y$ plane to fulfill the two first conditions. We may define $\nabla\cdot\vec{B}=\rho_m$, where $\rho_m$ is a uniform monopole magnetic volume density and $B_{HMW}=\rho_m\hat{\e}_z$.

Again choosing the symmetric gauge as
\begin{eqnarray}
	\vec{B}=\frac{\rho_m}{2}r\hat{\mathrm{e}}_r\;,
	\label{eqa:12}
\end{eqnarray}
and we obtain the following effective potential
\begin{eqnarray}
	\vec{A}_{AC}&=&\frac{\rho_m}{2}r\hat{\mathrm{e}}_\phi\;,
\end{eqnarray}
and rewrite the Eq.(\ref{eqa:9}) in the form
\begin{equation}
	H=\frac{1}{2m}\left[\vec{p}+\frac{m\omega}{2}r\hat{\mathrm{e}}_\phi\right]^2+\frac{\omega}{2}\;,
	\label{eqa:13}
\end{equation}
where
\begin{equation}
	\omega=\omega_{HMW}=\frac{d\rho_m}{m}\;
	\label{eqa:14}
\end{equation}
is the cyclotron frequency. Therefore, The Landau--He--McKellar--Wilkens energy levels are given by\cite{lrr:pla1}
\begin{equation}
	\mathcal{E}=\left(n+\frac{|\ell|+\ell-1}{2}+\frac{1}{2}\right)\omega_{HMW}\;,
	\label{eqa:15}
\end{equation}
where $n=0,1,2\dots$. So, we have energy levels for neutral electric polarized particles in the same way of Landau quantization for charged particles is a homogeneous magnetic field.

\section{Noncommutative Quantum Mechanics}\label{4}
In quantum mechanics, several problems have been investigated in the noncommutative space. Some interesting results are related to geometric phases \cite{1,2,3,4,5,mirza,kangepjc,lrr:pra}, and others effects that involve dynamics of dipoles \cite{cal}. The idea is to map the noncommutative on commutative space by replacing the coordinates $x^i$ and momenta $p^i$ by Hermitian operators $\hat{x}^i$ and $\hat{p}^i$ which obeys the relations
\begin{eqnarray}
	[\hat{x}^i,\hat{x}^j]&=&\I\theta^{ij}\;,\nonumber\\
	\label{eqd:1}[\hat{p}^i,\hat{p}^j]&=&0\;,\\
	\nonumber[\hat{x}^i,\hat{p}^j]&=&\I\delta^{ij}\;,
\end{eqnarray}
where $\theta^{ij}=\theta\epsilon^{ij}$ and $\epsilon^{ij}$ is the antisymmetric tensor. The time-independent Schr\"odinger equation in the noncommutative space may be written in the form
\begin{equation}
	H(x,p)\star\psi=\mathcal{E}\psi\;,
	\label{eqd:2}
\end{equation}
where $H(x,p)$ is the usual Hamiltonian and the Moyal product is defined by
\begin{equation}
	(f\star g)(x)=\exp\left(\frac{\I}{2}\theta^{ij}\partial_{x^i}\partial_{x^j}\right)f(x^i)g(x^j)\;.
	\label{eqd:3}
\end{equation}
Here $f$ and $g$ are arbitrary functions. In the noncommutative quantum mechanics, the Moyal product may be replaced by a Bopp shifts \cite{Bopp}, i.e., the $\star$-product may be changed into a ordinary product by replacing $H(x,p)$ on $H(\hat{x},\hat{p})$ as follows
\begin{equation}
	H(\hat{x}^i,\hat{p}^i)=H\left(x^i-\frac12\theta\epsilon^{ij}p^j,p^i\right)\;,
	\label{eqd:4}
\end{equation}
where $x^i$ and $p^i$ are the generalized position and momentum coordinates in the usual quantum mechanics. Therefore, the Eq.(\ref{eqd:4}) is defined on the commutative space and the effects due to noncommutativity may be calculated from the terms that contain the parameter $\theta$. Thus, we must change $x$ in the Schr\"odinger equation by a Bopp shifts
\begin{equation}
	\hat{x}^i\to x^i-\frac12\theta\epsilon^{ij}p^j\;.
	\label{eqd:4.1}
\end{equation}

Now we consider the case in which both space-space and momentum-momentum coordinates do not commute. The Bose--Einstein statistics in noncommutative quantum  mechanics requires this kind of formulation \cite{nair,zhang}. This is called phase-space noncommutativity. In this case the operators $x^i$ and $p^i$ obeys the commutation relations
\begin{eqnarray}
	[\hat{x}^i,\hat{x}^j]&=&\I\theta^{ij}\;,\nonumber\\
	\label{eqd:5}[\hat{p}^i,\hat{p}^j]&=&\I\bar{\theta}^{ij}\;,\\
	\nonumber[\hat{x}^i,\hat{p}^j]&=&\I\delta^{ij}\;,
\end{eqnarray}
where $\bar{\theta}^{ij}$ may be also a antisymmetric constant tensor, $\bar{\theta}^{ij}=\bar{\theta}\epsilon^{ij}$. Here, we also have the Bopp shifts
\begin{equation}
	H(\hat{x}^i,\hat{p}^i)=H\left(\lambda x^i-\frac{1}{2\lambda}\theta\epsilon^{ij}p^j,\lambda p^i+\frac{1}{2\lambda}\bar{\theta}\epsilon^{ij}x^j\right)\;,
	\label{eqd:6}
\end{equation}
where the constant $\lambda$ is a scaling factor. Thus, to map the noncommutative phase-space on commutative space we may change $\hat{x}$ and $\hat{p}$ by a Bopp shifts
\begin{eqnarray}
	\hat{x}^i&\to&\lambda x^i-\frac{1}{2\lambda}\theta\epsilon^{ij}p^j\;,\nonumber\\[-3mm]
	\label{eqd:7}\\[-3mm]
	\hat{p}^i&\to&\lambda p^i+\frac{1}{2\lambda}\bar{\theta}\epsilon^{ij}x^j\;.\nonumber
\end{eqnarray}

In the following sections, we investigate the analogs of Landau quantization in the noncommutative space and phase space. The standard Landau levels in the noncommutative space are studied by Horvathy\cite{hvarti2} and Gamboa et al. \cite{gamboa2}.

\section{Lan\-dau--A\-ha\-ro\-nov--Ca\-sher Le\-vels in the Non\-commutative Space}\label{5}
Now, in this section we analyze the Landau-Aharonov-Casher in the point of view of the non-commutative quantum mechanics.
In the case where $d$ and $\vec{B}$ are vanishes in the equation (\ref{eqa:1}), the Hamiltonian, for a magnetic dipole moment in the presence of a external electric field, is write in the form:
\begin{equation}
	H=-\frac{1}{2m}\left[\nabla-\I\mu\vec{A}_{AC}\right]^2-\frac{\mu}{2m}\vec{\nabla}\cdot\vec{E}\;,
	\label{eq:1}
\end{equation}
where
\begin{equation}
	\vec{A}_{AC}=\vec{n}\times\vec{E}\;,\qquad \vec{n}=\frac{\vec{\mu}}{|\vec{\mu}|}\;,
	\label{eq:2}
\end{equation}
and $\vec{n}$ is unitary and oriented in the dipole direction.

Now we choose the dipole orientation in $z$-axis, $\vec{n}=(0,0,1)$, and  the following electric field
\begin{eqnarray}
	\vec{E}&=&\frac{\rho}{2}(x,y,0)\;,\label{eq:3}
\end{eqnarray}
in this way, we obtain  the following effective potential
\begin{eqnarray}
	\vec{A}_{AC}&=&\frac{\rho}{2}(-y,x,0)\;,
\end{eqnarray}
and we rewrite the Eq.(\ref{eq:1}) in the form
\begin{equation}
	H=\frac{1}{2m}\left[\left(p_x+\frac{\mu\rho}{2}y\right)^2+\left(p_y-\frac{\mu\rho}{2}x\right)^2\right]-\frac{\mu\rho}{2m}\;.
	\label{eq:4}
\end{equation}

We can map the noncommutative space on the commutative space by a Bopp shifts, so the coordinates change
\begin{eqnarray}
	x&\to& x-\frac{\theta}{2}p_y\;,\nonumber\\[-3mm]
	&&\label{eq:5}\\[-3mm]
	y&\to& y+\frac{\theta}{2}p_x\;,\nonumber
\end{eqnarray}
and the Eq.(\ref{eq:4}) takes the form
\begin{eqnarray}
	H&=&\frac{1}{2m}\left[\left(\left(1+\frac{\mu\rho\theta}{4}\right)p_x+\frac{\mu\rho}{2}y\right)^2+ \left(\left(1+\frac{\mu\rho\theta}{4}\right)p_y-\frac{\mu\rho}{2}x\right)^2\right]-\frac{\mu\rho}{2m}\;.
\end{eqnarray}
We redefine the mass and frequency and we obtain the following Hamiltonian
\begin{eqnarray}
	H&=&\frac{1}{2\tilde{m}}\left[\left(p_x+\frac{\tilde{m}\tilde{\omega}}{2}y\right)^2+\left(p_y-\frac{\tilde{m}\tilde{\omega}}{2}x\right)^2\right]-\frac{\tilde{\omega}}{2}\left(1+\frac{\mu\rho\theta}{4}\right)^{-1}\;,
	\label{eq:6}
\end{eqnarray}
where
\begin{equation}
	\tilde{m}=\dfrac{m}{\left(1+\dfrac{\mu\rho\theta}{4}\right)^{2}}\;,\qquad \tilde{\omega}=\dfrac{\mu\rho}{\tilde{m}\left(1+\dfrac{\mu\rho\theta}{4}\right)}\;.
	\label{eq:7}
\end{equation}
The noncommutative contributions redefines the mass and cyclotron frequency of the dipole.

To solve the Schr\"odinger equation, we may make use the cylindrical symmetry of the system and rewrite the Hamiltonian (\ref{eq:6}) as
\begin{equation}
	H=\frac{1}{2\tilde{m}}\left[\vec{p}-\frac{\tilde{m}\tilde{\omega}}{2}r\hat{\mathrm{e}}_\phi\right]^2-\frac{\tilde{\omega}}{2}\left(1+\frac{\mu\rho\theta}{4}\right)^{-1}\;.
	\label{eq:8}
\end{equation}
The Schr\"odinger equation in cylindrical coordinates is written in the form
\begin{equation}
	-\frac{1}{2\tilde{m}}\left[\frac{1}{r}\frac{\partial}{\partial r}\left(r\frac{\partial\psi}{\partial r}\right)+\frac{1}{r^2}\frac{\partial^2\psi}{\partial\phi^2}\right]-\frac{\I\tilde{\omega}}{2}\frac{\partial\psi}{\partial\phi}+\frac{\tilde{m}\tilde{\omega}^2}{8}r^2\psi-\frac{\tilde{\omega}}{2}\left(1+\frac{\mu\rho\theta}{4}\right)^{-1}\psi=\mathcal{E}\psi\;.
	\label{eq:9}
\end{equation}

We use the following ansatz to the solution of Eq.(\ref{eq:9})
\begin{equation}
	\psi=\e^{\I\ell\phi}R(r)\;,
	\label{eq:10}
\end{equation}
where $\ell$ is an integer number. Thus, we can rewrite the Eq.(\ref{eq:9}) as
\begin{equation}
	\frac{1}{2\tilde{m}}\left(R''+\frac{1}{r}R'-\frac{\tilde{m}^2}{r^2}R\right)+ \left(\mathcal{E}-\frac{\tilde{m}\tilde{\omega}^2}{8}r^2+\frac{\ell\tilde{\omega}}{2}+\frac{\tilde{\omega}}{2}\left(1+\frac{\mu\rho\theta}{4}\right)^{-1}\right)R=0\;,
	\label{eq:11}
\end{equation}
and using the following change of variables
\begin{equation}
	\xi=\frac{\tilde{m}\tilde{\omega}}{2}r^2\;,
	\label{eq:12}
\end{equation}
the radial Schr\"odinger equation is rewritten in the form
\begin{equation}
	\xi R''+R'+\left(-\frac{\xi}{4}+\beta-\frac{\ell^2}{4\xi}\right)R=0\;,
	\label{eq:13}
\end{equation}
where
\begin{equation}
	\beta=\frac{\mathcal{E}}{\tilde{\omega}}+\dfrac{\ell}{2}+\frac{1}{2}\left(1+\dfrac{\mu\rho\theta}{4}\right)^{-1}\;.
	\label{eq:14}
\end{equation}
Studying the asymptotic limit of the solutions in the Eq.(\ref{eq:13}), we may write the solution in the form
\begin{equation}
	R(\xi)=\e^{-\xi/2}\xi^{|\ell|/2}\zeta(\xi)\;.
	\label{eq:15}
\end{equation}
So, the hypergeometric equation that is satisfied by the function $\zeta(\xi)$ is given by
\begin{equation}
	\zeta=F\left[-\left(\beta-\frac{|\ell|+1}{2}\right),|\ell|+1,\xi\right]\;.
	\label{eq:16}
\end{equation}
The condition to Eq.(\ref{eq:16}) be finite is that the first term in hypergeometric must be a non-positive integer. Then, the noncommutative Landau--Aharonov--Casher energy levels are given by
\begin{equation}
	\mathcal{E}=\left(n+\frac{|\ell|-\ell+1}{2}\right)\tilde{\omega}-\dfrac{1}{2}\left(\dfrac{1+\mu\rho\theta}{4}\right)^{-1}\tilde{\omega}\;,
	\label{eq:17}
\end{equation} 
where $n=0,1,2,\dots$. The radial energy eigenfunctions are given by
\begin{equation}
	R_{n,\ell}(r)=\frac{1}{\tilde{a}^{|\ell|+1}}\left[\frac{(|\ell|+n)!}{2^{|\ell|}n!|\ell|!^2}\right]\e^{-\frac{r^2}{4\tilde{a}^2}}r^{|\ell|}F\left[-n,|\ell|+1,\frac{r^2}{2\tilde{a}^2}\right]\;,
	\label{eq:18}
\end{equation}
where
\begin{equation}
	\tilde{a}=\sqrt{\frac{1}{\tilde{m}\tilde{\omega}}}\;
	\label{eq:19}
\end{equation}
is the new magnetic length redefined by the noncommutativy of space coordinates. 

If we make $\theta$ vanishes in Eq.(\ref{eq:17}), we retrieve the original result in Eq.(\ref{eqa:8}). It's easy to see that the noncommutative corrections shifts up the energy levels and reduces the magnetic length.

\section{Lan\-dau--A\-ha\-ro\-nov--Ca\-sher in the Noncommutative Phase-Space}\label{6}
We analyze in this section the Landau-Aharonov-Casher problem in nocomutative phase space using the description adopted in section \ref{4}.
To map the noncommutative phase-space in the commutative space, we have the following Bopp shifts
\begin{eqnarray}
	x\to\lambda x-\frac{\theta}{2\lambda}p_y\;,\qquad p_x\to\lambda p_x+\frac{\bar{\theta}}{2\lambda}y\;,\nonumber\\[-3mm]
	\label{eq:20}\\[-3mm]
	y\to\lambda y+\frac{\theta}{2\lambda}p_x\;,\qquad p_y\to\lambda p_y-\frac{\bar{\theta}}{2\lambda}x\;,\nonumber		
\end{eqnarray}
where the scale factor $\lambda$ is a arbitrary constant parameter. So, the Eq.(\ref{eq:4}) becomes
\begin{eqnarray}
H&=&\frac{1}{2m}\left[\left(\left(\lambda+\frac{\mu\rho\theta}{4\lambda}\right)p_x+\frac12\left(\mu\rho\lambda+\frac{\bar{\theta}}{\lambda}\right)y\right)^2\right.\nonumber\\
&&\qquad+\left.\left(\left(\lambda+\frac{\mu\rho\theta}{4\lambda}\right)p_y- \frac12\left(\mu\rho\lambda+\frac{\bar{\theta}}{\lambda}\right)x\right)^2\right]-\frac{\mu\rho}{2m}\;.
\end{eqnarray}
Perforating a rescaling in the mass and frequency we obtain
\begin{eqnarray}
	H&=&\frac{1}{2\tilde{m}}\left[\left(p_x+\frac{\tilde{m}\tilde{\omega}}{2}y\right)^2+\left(p_y-\frac{\tilde{m}\tilde{\omega}}{2}x\right)^2\right]\nonumber\\[-2mm]
&&\label{eq:21}\\[-2mm]
&&-\frac{\tilde{\omega}}{2\lambda}\left(\lambda+\dfrac{\mu\rho\theta}{4\lambda}\right)^{-1}+ \frac{\bar{\theta}}{2\tilde{m}\lambda^2}\left(\lambda+\dfrac{\mu\rho\theta}{4\lambda}\right)^{-2}\;,\nonumber
\end{eqnarray}
where
\begin{equation}
\tilde{m}=\dfrac{m}{\left(\lambda+\dfrac{\mu\rho\theta}{4\lambda}\right)^2}\;,\qquad\tilde{\omega}=\dfrac{\left(\mu\rho\lambda+\dfrac{\bar{\theta}}{\lambda}\right)}{\tilde{m}\left(\lambda+\dfrac{\mu\rho\theta}{4\lambda}\right)}\;.
	\label{eq:22}
\end{equation}
Here, we redefined the mass and cyclotron frequency in terms of the noncommutativity parameters, $\theta$ and $\bar\theta$.
Making use of the cylindrical symmetry of the problem, we may rewrite the Hamiltonian (\ref{eq:21}) in the form
\begin{equation}
	H=\frac{1}{2\tilde{m}}\left[\vec{p}-\frac{\tilde{m}\tilde{\omega}}{2}r\hat{\mathrm{e}}_\phi\right]^2-\frac{\tilde{\omega}}{2\lambda}\left(\lambda+\dfrac{\mu\rho\theta}{4\lambda}\right)^{-1}+ \frac{\bar{\theta}}{2\tilde{m}\lambda^2}\left(\lambda+\dfrac{\mu\rho\theta}{4\lambda}\right)^{-2}\;.
	\label{eq:23}
\end{equation}

The Schr\"odinger equation in given by
\begin{eqnarray}
	-\frac{1}{2\tilde{m}}\left[\frac{1}{r}\frac{\partial}{\partial r}\left(r\frac{\partial\psi}{\partial r}\right)+\frac{1}{r^2}\frac{\partial^2\psi}{\partial\phi^2}\right]-\frac{\I\tilde{\omega}}{2}\frac{\partial\psi}{\partial\phi}+\frac{\tilde{m}\tilde{\omega}^2}{8}r^2\psi\nonumber\\-\frac{\tilde{\omega}}{2\lambda}\left(\lambda+\dfrac{\mu\rho\theta}{4\lambda}\right)^{-1}\psi+ \frac{\bar{\theta}}{2\tilde{m}\lambda^2}\left(\lambda+\dfrac{\mu\rho\theta}{4\lambda}\right)^{-2}\psi=\mathcal{E}\psi\;.
	\label{eq:24}
\end{eqnarray}
We use the following ansatz to the solution of Eq.(\ref{eq:24})
\begin{equation}
	\psi=\e^{\I\ell\phi}R(r)\;,
	\label{eq:25}
\end{equation}
and write the radial Sch\"odinger equation as
\begin{equation}
	\xi R''+R'+\left(-\frac{\xi}{4}+\beta-\frac{\ell^2}{4\xi}\right)R=0\;,
	\label{eq:26}
\end{equation}
where we used the change in variables
\begin{equation}
	\xi=\frac{\tilde{m}\tilde{\omega}}{2}r^2\;,
	\label{eq:27}
\end{equation}
and
\begin{equation}
	\beta=\frac{\mathcal{E}}{\tilde{\omega}}+\frac{\ell}{2}+\frac{1}{2\lambda}\left(\lambda+\dfrac{\mu\rho\theta}{4\lambda}\right)^{-1}- \frac{\bar{\theta}}{2\tilde{m}\tilde{\omega}\lambda^2}\left(\lambda+\dfrac{\mu\rho\theta}{4\lambda}\right)^{-2}\;.
	\label{eq:28}
\end{equation}
Studying the asymptotic limit of the solutions in the Eq.(\ref{eq:26}), we may write the solution in the form
\begin{equation}
	R(\xi)=\e^{-\xi/2}\xi^{|\ell|/2}\zeta(\xi)\;.
	\label{eq:29}
\end{equation}
So, the hypergeometric equation that is satisfied by the function $\zeta(\xi)$ is given by
\begin{equation}
	\zeta=F\left[-\left(\beta-\frac{|\ell|+1}{2}\right),|\ell+1|,\xi\right]\;.
	\label{eq:30}
\end{equation}
The condition to Eq.(\ref{eq:30}) be finite is that the first term in hypergeometric must be a non-positive integer. Then, the phase space noncommutative Landau--Aharonov--Casher energy levels are given by
\begin{equation}
	\mathcal{E}=\left(n+\frac{|\ell|-\ell+1}{2}\right)\tilde{\omega}-\frac{1}{2\lambda}\left(\lambda+\dfrac{\mu\rho\theta}{4\lambda}\right)^{-1}\tilde{\omega}+ \frac{\bar{\theta}}{2\tilde{m}\lambda^2}\left(\lambda+\dfrac{\mu\rho\theta}{4\lambda}\right)^{-2}\;,
	\label{eq:31}
\end{equation} 
where $n=0,1,2,\dots$. The radial energy eigenfunctions are given by
\begin{equation}
	R_{n,\ell}(r)=\frac{1}{\tilde{a}^{|\ell|+1}}\left[\frac{(|\ell|+n)!}{2^{|\ell|}n!|\ell|!^2}\right]\e^{-\frac{r^2}{4\tilde{a}^2}}r^{|\ell|}F\left[-n,|\ell|+1,\frac{r^2}{2\tilde{a}^2}\right]\;,
	\label{eq:32}
\end{equation}
where
\begin{equation}
	\tilde{a}=\sqrt{\frac{1}{\tilde{m}\tilde{\omega}}}\;
	\label{eq:33}
\end{equation}
is the redefined magnetic length. It's easy to see that we retrieve the original effect if we make the noncommutative parameters $\theta$ and $\bar\theta$ vanish.

\section{Lan\-dau--He--McKellar--Wilkens Le\-vels in the Non\-com\-mu\-ta\-ti\-ve Space}\label{7}
In the same way of the Landau--Aharonov--Casher levels, in the case that $\mu$ and $\vec{E}$ vanishes in the equation (\ref{eqa:1}), we write the Hamiltonian for a electric dipole moment in the presence of a external magnetic field in the form:
\begin{equation}
	H=-\frac{1}{2m}\left[\nabla+\I d\vec{A_{HMW}}\right]^2+\frac{d}{2m}\vec{\nabla}\cdot\vec{B}\;,
	\label{eqb:1}
\end{equation}
where
\begin{equation}
	\vec{A}_{HMW}=\vec{n}\times\vec{B}\;,\qquad \vec{n}=\frac{\vec{d}}{|\vec{d}|}\;,
	\label{eqb:2}
\end{equation}
and $\vec{n}$ is unitary and oriented in the dipole direction.

Now we choose the dipole orientation in $z$-axis, $\vec{n}=(0,0,1)$, and in the symmetric gauge
\begin{equation}
	\vec{B}=\frac{\rho_m}{2}(x,y,0)\;,\label{eqb:3}
\end{equation}
and we obtain the effective vector potential
\begin{equation}
	\vec{A}_{HMW}=\frac{\rho_m}{2}(-y,x,0)\;,
\end{equation}
where $\rho_m$ is the magnetic monopole charge density. We rewrite the Eq.(\ref{eqb:1}) in the form
\begin{equation}
	H=\frac{1}{2m}\left[\left(p_x-\frac{d\rho_m}{2}y\right)^2+\left(p_y+\frac{d\rho_m}{2}x\right)^2\right]+\frac{d\rho_m}{2m}\;.
	\label{eqb:4}
\end{equation}

We can map the noncommutative space in the commutative space by a Bopp shifts, so the coordinates change
\begin{eqnarray}
	x&\to& x-\frac{\theta}{2}p_y\;,\nonumber\\[-2mm]
	&&\label{eqb:5}\\[-2mm]
	y&\to& y+\frac{\theta}{2}p_x\;,\nonumber
\end{eqnarray}
and the Eq.(\ref{eq:4}) takes the form
\begin{equation}
	H=\frac{1}{2m}\left[\left(\left(1+\frac{d\rho_m\theta}{4}\right)p_x-\frac{d\rho_m}{2}y\right)^2+ \left(\left(1+\frac{d\rho_m\theta}{4}\right)p_y+\frac{d\rho_m}{2}x\right)^2\right]+\frac{d\rho_m}{2m}\;.\label{eqb:6.1}
\end{equation}
Perforating a rescaling in the mass and frequency we obtain
\begin{equation}
H=\frac{1}{2\tilde{m}}\left[\left(p_x-\frac{\tilde{m}\tilde{\omega}}{2}y\right)^2+\left(p_y+\frac{\tilde{m}\tilde{\omega}}{2}x\right)^2\right]+\frac{\tilde{\omega}}{2}\left(1+\frac{d\rho_m\theta}{4}\right)^{-1}\;,
	\label{eqb:6}
\end{equation}
where
\begin{equation}
	\tilde{m}=\dfrac{m}{\left(1+\dfrac{d\rho_m\theta}{4}\right)^{2}}\;,\qquad \tilde{\omega}=\dfrac{d\rho_m}{\tilde{m}\left(1+\dfrac{d\rho_m\theta}{4}\right)}\;.
	\label{eqb:7}
\end{equation}
The noncommutative contributions redefines the mass and cyclotron frequency of the dipole.

To solve the Schr\"odinger equation, we may make use the cylindrical symmetry of the system and rewrite the Hamiltonian (\ref{eqb:6})
\begin{equation}
	H=\frac{1}{2\tilde{m}}\left[\vec{p}+\frac{\tilde{m}\tilde{\omega}}{2}r\hat{\mathrm{e}}_\phi\right]^2+\frac{\tilde{\omega}}{2}\left(1+\frac{d\rho_m\theta}{4}\right)^{-1}\;.
	\label{eqb:8}
\end{equation}
The Schr\"odinger equation in cylindrical coordinates is written in the form
\begin{equation}
	-\frac{1}{2\tilde{m}}\left[\frac{1}{r}\frac{\partial}{\partial r}\left(r\frac{\partial\psi}{\partial r}\right)+\frac{1}{r^2}\frac{\partial^2\psi}{\partial\phi^2}\right]-\frac{\I\tilde{\omega}}{2}\frac{\partial\psi}{\partial\phi}+\frac{\tilde{m}\tilde{\omega}^2}{8}r^2\psi+\frac{\tilde{\omega}}{2}\left(1+\frac{d\rho_m\theta}{4}\right)^{-1}\psi=\mathcal{E}\psi\;.
	\label{eqb:9}
\end{equation}

We use the following ansatz to the solution of Eq.(\ref{eqb:9})
\begin{equation}
	\psi=\e^{\I\ell\phi}R(r)\;,
	\label{eqb:10}
\end{equation}
where $\ell$ is an integer number. Hence, from the Eq.(\ref{eqb:9}) is write as
\begin{equation}
	\frac{1}{2\tilde{m}}\left(R''+\frac{1}{r}R'-\frac{\tilde{m}^2}{r^2}R\right)+ \left(\mathcal{E}-\frac{\tilde{m}\tilde{\omega}^2}{8}r^2+\frac{\ell\tilde{\omega}}{2}-\frac{\tilde{\omega}}{2}\left(1+\frac{d\rho_m\theta}{4}\right)^{-1}\right)R=0\;,
	\label{eqb:11}
\end{equation}
and using the following change of variables
\begin{equation}
	\xi=\frac{\tilde{m}\tilde{\omega}}{2}r^2\;,
	\label{eqb:12}
\end{equation}
the radial Schr\"odinger equation is rewritten in the form
\begin{equation}
	\xi R''+R'+\left(-\frac{\xi}{4}+\beta-\frac{\ell^2}{4\xi}\right)R=0\;,
	\label{eqb:13}
\end{equation}
where
\begin{equation}
	\beta=\frac{\mathcal{E}}{\tilde{\omega}}+\dfrac{\ell}{2}-\frac{1}{2}\left(1+\dfrac{d\rho_m\theta}{4}\right)^{-1}\;.
	\label{eqb:14}
\end{equation}
Studying the asymptotic limit of the solutions in the Eq.(\ref{eqb:13}), we may write the solution in the form
\begin{equation}
	R(\xi)=\e^{-\xi/2}\xi^{|\ell|/2}\zeta(\xi)\;.
	\label{eqb:15}
\end{equation}
So, the hypergeometric equation that is satisfied by the function $\zeta(\xi)$ is given by
\begin{equation}
	\zeta=F\left[-\left(\beta-\frac{|\ell|+1}{2}\right),|\ell|+1,\xi\right]\;.
	\label{eqb:16}
\end{equation}
The condition to Eq.(\ref{eqb:16}) be finite is that the first term in hypergeometric must be a non-positive integer. Then, the noncommutative Landau--He--McKellar--Wilkens energy levels are given by
\begin{equation}
	\mathcal{E}=\left(n+\frac{|\ell|+\ell-1}{2}\right)\tilde{\omega}+\dfrac{1}{2}\left(\dfrac{1+d\rho_m\theta}{4}\right)^{-1}\tilde{\omega}\;,
	\label{eqb:17}
\end{equation} 
where $n=0,1,2,\dots$. The radial energy eigenfunctions are given by
\begin{equation}
	R_{n,\ell}(r)=\frac{1}{\tilde{a}^{|\ell|+1}}\left[\frac{(|\ell|+n)!}{2^{|\ell|}n!|\ell|!^2}\right]\e^{-\frac{r^2}{4\tilde{a}^2}}r^{|\ell|}F\left[-n,|\ell|+1,\frac{r^2}{2\tilde{a}^2}\right]\;,
	\label{eqb:18}
\end{equation}
where
\begin{equation}
	\tilde{a}=\sqrt{\frac{1}{\tilde{m}\tilde{\omega}}}\;
	\label{eqb:19}
\end{equation}
is the magnetic length redefined by the noncommutative parameters.

If we make $\theta$ zero in Eq.(\ref{eqb:17}), we retrieve the original effect in Eq.(\ref{eqa:15}). It's easy to see that the noncommutative corrections shifts up the energy levels and reduces the magnetic length.

\section{Lan\-dau--He--McKellar--Wilkens Levels in the Non\-com\-mu\-ta\-ti\-ve Phase-Space}\label{8}
To map the noncommutative phase-space in the commutative space, we have the following Bopp shifts
\begin{eqnarray}
	x\to\lambda x-\frac{\theta}{2\lambda}p_y\;,\qquad p_x\to\lambda p_x+\frac{\bar{\theta}}{2\lambda}y\;,\nonumber\\[-3mm]
	\label{eqb:20}\\[-3mm]
	y\to\lambda y+\frac{\theta}{2\lambda}p_x\;,\qquad p_y\to\lambda p_y-\frac{\bar{\theta}}{2\lambda}x\;,\nonumber		
\end{eqnarray}
where the scale factor $\lambda$ is a arbitrary constant parameter. So, the Eq.(\ref{eqb:4}) becomes
\begin{eqnarray}
	H&=&\frac{1}{2m}\left[\left(\left(\lambda+\frac{d\rho_m\theta}{4\lambda}\right)p_x-\frac12\left(d\rho_m\lambda+\frac{\bar{\theta}}{\lambda}\right)y\right)^2\right.\nonumber\\
&&\qquad+\left.\left(\left(\lambda+\frac{d\rho_m\theta}{4\lambda}\right)p_y+ \frac12\left(d\rho_m\lambda+\frac{\bar{\theta}}{\lambda}\right)x\right)^2\right]-\frac{d\rho_m}{2m}
\end{eqnarray}
redefining the frequency and mass we have 
\begin{eqnarray}
	H&=&\frac{1}{2\tilde{m}}\left[\left(p_x-\frac{\tilde{m}\tilde{\omega}}{2}y\right)^2+\left(p_y+\frac{\tilde{m}\tilde{\omega}}{2}x\right)^2\right]\nonumber\\[-2mm]
&&\label{eqb:21}\\[-2mm]
&&+\frac{\tilde{\omega}}{2\lambda}\left(\lambda+\dfrac{d\rho_m\theta}{4\lambda}\right)^{-1}- \frac{\bar{\theta}}{2\tilde{m}\lambda^2}\left(\lambda+\dfrac{d\rho_m\theta}{4\lambda}\right)^{-2}\;,\nonumber
\end{eqnarray}
where
\begin{equation}
	\tilde{m}=\dfrac{m}{\left(\lambda+\dfrac{d\rho_m\theta}{4\lambda}\right)^2}\;,\qquad\tilde{\omega}=\dfrac{\left(d\rho_m\lambda+\dfrac{\bar{\theta}}{\lambda}\right)}{\tilde{m}\left(\lambda+\dfrac{d\rho_m\theta}{4\lambda}\right)}\;.
	\label{eqb:22}
\end{equation}
Making use of the cylindrical symmetry of the problem, we may rewrite the Hamiltonian (\ref{eqb:21}) in the form
\begin{equation}
	H=\frac{1}{2\tilde{m}}\left[\vec{p}+\frac{\tilde{m}\tilde{\omega}}{2}r\hat{\mathrm{e}}_\phi\right]^2+\frac{\tilde{\omega}}{2\lambda}\left(\lambda+\dfrac{d\rho_m\theta}{4\lambda}\right)^{-1}-\frac{\bar{\theta}}{2\tilde{m}\lambda^2}\left(\lambda+\dfrac{d\rho_m\theta}{4\lambda}\right)^{-2}\;.
	\label{eqb:23}
\end{equation}

The Schr\"odinger equation in given by
\begin{eqnarray}
	-\frac{1}{2\tilde{m}}\left[\frac{1}{r}\frac{\partial}{\partial r}\left(r\frac{\partial\psi}{\partial r}\right)+\frac{1}{r^2}\frac{\partial^2\psi}{\partial\phi^2}\right]-\frac{\I\tilde{\omega}}{2}\frac{\partial\psi}{\partial\phi}+\frac{\tilde{m}\tilde{\omega}^2}{8}r^2\psi\nonumber\\+\frac{\tilde{\omega}}{2\lambda}\left(\lambda+\dfrac{d\rho_m\theta}{4\lambda}\right)^{-1}\psi- \frac{\bar{\theta}}{2\tilde{m}\lambda^2}\left(\lambda+\dfrac{d\rho_m\theta}{4\lambda}\right)^{-2}\psi=\mathcal{E}\psi\;.
	\label{eqb:24}
\end{eqnarray}
We use the following ansatz to the solution of Eq.(\ref{eqb:24})
\begin{equation}
	\psi=\e^{\I\ell\phi}R(r)\;,
	\label{eqb:25}
\end{equation}
and the radial Sch\"odinger equation is
\begin{equation}
	\xi R''+R'+\left(-\frac{\xi}{4}+\beta-\frac{\ell^2}{4\xi}\right)R=0\;,
	\label{eqb:26}
\end{equation}
where we used the change in variables
\begin{equation}
	\xi=\frac{\tilde{m}\tilde{\omega}}{2}r^2\;,
	\label{eqb:27}
\end{equation}
and
\begin{equation}
	\beta=\frac{\mathcal{E}}{\tilde{\omega}}+\frac{\ell}{2}-\frac{1}{2\lambda}\left(\lambda+\dfrac{d\rho_m\theta}{4\lambda}\right)^{-1}+ \frac{\bar{\theta}}{2\tilde{m}\tilde{\omega}\lambda^2}\left(\lambda+\dfrac{d\rho_m\theta}{4\lambda}\right)^{-2}\;.
	\label{eqb:28}
\end{equation}
Studying the asymptotic limit of the solutions in the Eq.(\ref{eqb:26}), we may write the solution in the form
\begin{equation}
	R(\xi)=\e^{-\xi/2}\xi^{|\ell|/2}\zeta(\xi)\;.
	\label{eqb:29}
\end{equation}
So, the hypergeometric equation that is satisfied by the function $\zeta(\xi)$ is given by
\begin{equation}
	\zeta=F\left[-\left(\beta-\frac{|\ell|+1}{2}\right),|\ell|+1,\xi\right]\;.
	\label{eqb:30}
\end{equation}
The condition to Eq.(\ref{eqb:30}) be finite is that the first term in hypergeometric must be a non-positive integer. Then, the phase space noncommutative Landau--He--McKellar--Wilkens energy levels are given by
\begin{equation}
	\mathcal{E}=\left(n+\frac{|\ell|+\ell-1}{2}\right)\tilde{\omega}+\frac{1}{2\lambda}\left(\lambda+\dfrac{d\rho\theta}{4\lambda}\right)^{-1}\tilde{\omega}- \frac{\bar{\theta}}{2\tilde{m}\lambda^2}\left(\lambda+\dfrac{d\rho\theta}{4\lambda}\right)^{-2}\;,
	\label{eqb:31}
\end{equation} 
where $n=0,1,2,\dots$. The radial energy eigenfunctions are given by
\begin{equation}
	R_{n,\ell}(r)=\frac{1}{\tilde{a}^{|\ell|+1}}\left[\frac{(|\ell|+n)!}{2^{|\ell|}n!|\ell|!^2}\right]\e^{-\frac{r^2}{4\tilde{a}^2}}r^{|\ell|}F\left[-n,|\ell|+1,\frac{r^2}{2\tilde{a}^2}\right]\;,
	\label{eqb:32}
\end{equation}
where
\begin{equation}
	\tilde{a}=\sqrt{\frac{1}{\tilde{m}\tilde{\omega}}}\;
	\label{eqb:33}
\end{equation}
is the new magnetic length redefined by the noncommutative parameters. It's easy to see that we retrieve the original effect if we make the noncommutative parameters $\theta$ and $\bar\theta$ vanish.

\section{Concluding Remarks}\label{9}
We studied the Landau analogs energy levels for neutral particles, that possesses nonzero magnetic and electric dipole moments, in the presence of homogeneous electric and magnetic external fields in the context of noncommutative quantum mechanics. We analyzed the Landau-Aharonov-Casher  and Landau-He-McKellar-Wilkens quantization to magnetic and electric dipole respectively. In both cases, we calculate the corrections in analog Landau energy levels due to the noncommutativity in the space and phase space coordinates. We also found the corrections to mass and cyclotron frequency in the noncummutative space and phase space, as well as the influence of noncommutativity in the energy levels, radial wave functions and magnetic length. Also, e verified that when we take the limit $\theta \to 0$ we retrieve the commutative result. 

{\bf Acknowledgments.} This work was partially supported by Funda\c c\~ao de Amparo \`a Pesquisa do Estado de S\~ao
Paulo (FAPESP), Conselho Nacional de Desenvolvimento Cient\'{\i}fico e Tecnol\'{o}gico (CNPq) and CAPES/PROCAD.


\begin{thebibliography}{99}
\bibliographystyle{unsrt}

\bibitem{aha} Y. Aharonov  and D. Bohm, Phys. Rev. {\bf 115}, 485 (1959).
\bibitem{prl:cham} R. G. Chambers, Phys. Rev. Lett. {\bf 5}, 3 (1960).
\bibitem{pes1} M.  Peshkin  and A. Tonomura, {\it The Aharonov-Bohm Effect} (Springer-Verlag, Berlin, 1989).
\bibitem{cas} Y. Aharonov and A. Casher, Phys. Rev. Lett. {\bf 53}, 319, (1984).
\bibitem{cim} A. Cimmino et al., Phys. Rev. Lett. {\bf 63}, 380 (1989).
\bibitem{san} K. Sangster et al., Phys. Rev. Lett. {\bf 71}, 3641 (1993).
\bibitem{mac} X. -G. He and B. H. J. McKellar, Phys. Rev. A {\bf 47}, 3424 (1993).
\bibitem{wil} M. Wilkens, Phys. Rev. Lett. {\bf 72}, 5 (1994).
\bibitem{wei} H. Wei, R. Han and X. Wei, Phys. Rev. Lett. {\bf 75}, 2071 (1995).
\bibitem{franson} J. P. Dowling, C. P. Willian and J. D. Franson, Phys. Rev. Lett. {\bf 83}, 2486 (1999).
\bibitem{furtadodual} Claudio Furtado and G. Duarte, Physica Scripta {\bf71},7(2005)
\bibitem{landau} L. D. Landau, Z. Phys. {\bf 64}, 629 (1930).
\bibitem{prange} R. E. Prange, S. M. Girvin (Eds.), The Quantum Hall Effect, Springer--Verlag, New York, 1990.
\bibitem{comtet} A. Comtet, Ann. Phys. (N.Y.) {\bf 173}, 185 (1987); C. Grosche, Ann. Phys. (N.Y.) {\bf 187}, 110 (1988).
\bibitem{dunne} G. V. dunne, Ann. Phys. (N.Y.) {\bf 215}, 233 (1992).
\bibitem{paredes1} B. Paredes, P. Fedichev, J. I. Cirac, P. Zoller, Phys. Rev. Lett. {\bf 87}, 010402 (2001).
\bibitem{paredes2} B. Paredes, P. Zoller, J. I. Cirac, Solid State Commun. {\bf 127}, 155 (2003).
\bibitem{ericsson} M. Ericsson and E. Sj\"oqvist, Phys. Rev A, {\bf 65}, 013607 (2001)
\bibitem{lrr:pla1} L. R. Ribeiro, C. Furtado and J. R. Nascimento, Phys. Lett. A {\bf 348}, 135 (2006).
\bibitem{lrr:pla2} C. Furtado, J. R. Nascimento and L. R. Ribeiro, Phys. Lett. A {\bf 358}, 336 (2006).
\bibitem{sei} N. Seiberg and E. Witten, JHEP {\bf 9909}, 032 (1999).
\bibitem{sei2} N. Seiberg, L. Susskind and N. Toumbas, JHEP {\bf 0006}, 044 (2000).
\bibitem{nekra} M. R. Douglas, N. A. Nekrasov, Rev. Mod. Phys. {\bf 73}, 977 (2001).
\bibitem{moffat} J. W. Moffat, Phys. Lett. B {\bf 493}, 142 (2000).
\bibitem{con} A. Connes, M. R. Douglas, A Schwarz, JHEP {\bf 9802}, 003 (1998).
\bibitem{sus} L. Susskind, e-print arXiv:hep-th/0101029v3.
\bibitem{jela} O. F. Dayi and A. Jellal, J. Math. Phys. {\bf 43}, 4592 (2002).
\bibitem{basu} B. Basu and  Subir Ghosh  Phys.Lett. A {\bf 346} 133 (2005).
\bibitem{hovarti}C.~Duval and P. A. Horvathy,  Phys. Lett. {\bf B 479}, 284 (2000)
\bibitem{1} M. Chaichian, M. M. Sheikh-Jabbari and A. Tureanu, Phys. Rev. Lett. {\bf 86}, 2716 (2001).
\bibitem{2} M. Chaichian, A. Demichev, P. Presnajder, M. M Sheikh-Jbbari and A. Tureanu, Nucl. Phys. B {\bf  611}, 383 (2001).
\bibitem{3} M. Chaichian, P. Presnajder, M. M Sheikh-Jabbari and A. Tureanu, Phys. Lett. {\bf B 527}, 149 (2002)
\bibitem{4} H. Falomir, J. Gamboa, M. Loeve, F. Mendez and J. C. Rojas, Phys. Rev. D {\bf 66}, 045018 (2002).
\bibitem{5} K. Li and S. Dulat, Eur. Phys. J. C {\bf 46} 825 (2006).
\bibitem{mirza}B. Mirza and M. Zarei, Eur. Phys. J. C {\bf 32} (2004).
\bibitem{kangepjc} K. Li and J. Wang, Eur. Phys. J. C {\bf 50} 1007 (2007).
\bibitem{alavi} S. A. Alavi, Physica Scripta T {\bf 7} 366 (2003).
\bibitem{ghos}B. Basu, Subir Ghosh, S. Dhar, Europhys.Lett.76:395,2006.
\bibitem{hvarti2} P. A. Horvathy, Ann. Phys. (N. Y.) {\bf 299}, 128-140 (2002).
\bibitem{hovarti3}P. A. Horvathy and M. S. Plyushchay. Nucl. Phys. {\bf B 714} 269 (2005).
\bibitem{gamboa2} J. Gamboa, M. Loewe, F. Mendez and J. C. Rojas, Mod. Phys. Lett. A {\bf 16}, 2075 (2001).
\bibitem{cal}X Calmet and M. Selvaggi Phys. Rev. D {\bf 74} 037901,(2006).
\bibitem{prlan} J. Anandan, Phys. Rev. Lett. {\bf 85}, 1354 (2000).
\bibitem{lrr:pra} E. Passos, L. R. Ribeiro, C. Furtado and J. R. Nascimento, Phys. Rev. A {\bf 76}, 012113 (2007).


\bibitem{Bopp} T. Curtright, D. Fairlie and C. Zachos, Phys. Rev. D {\bf 58}, 025002 (1998).

\bibitem{nair} V. P. Nair  and A. P. Polychronakos, Phys. Lett B {\bf 505}, 267 (2001).
\bibitem{zhang} J.-Z. Zhang, Phys Lett. B {\bf  584}, 204 (2004).


\end{thebibliography}
\end{document}